\documentclass[twocolumn,showpacs,showkeys,floatfix,preprintnumbers,amsmath,amssymb]{revtex4}

\usepackage{graphics}
\usepackage{graphicx}
\usepackage{amssymb}
\usepackage{hyperref}
\usepackage{color}

\begin{document}

\bibliographystyle{apsrev}

\title{Vortex phase diagram and temperature-dependent second-peak effect in overdoped Bi$_{2}$Sr$_{2}$CuO$_{6 +
\delta}$ crystals}

\author{A. Piriou $^{1}$} \email{Alexandre.Piriou@unige.ch}
\author{E. Giannini$^{1}$}
\author{Y. Fasano$^{1}$\footnote{Present address:  Instituto Balseiro and Centro
At\'omico Bariloche, Comisi\' on Nacional de Energ\'{i}a
At\'omica, Avenida Bustillo 9500, 8400 Bariloche, Argentina } }
\author{C. Senatore$^{1}$}
\author{\O. Fischer$^{1}$}

\affiliation{%
$^{1}$ D\'epartement de Physique de la Mati\`ere Condens\' ee,
Universit\'e de Gen\`eve, 24 Quai Ernest-Ansermet, 1211 Geneva,
Switzerland
}%

\date{\today}

\begin{abstract}

We study the vortex phase diagram of the single-layer
Bi$_{2}$Sr$_{2}$CuO$_{6+\delta}$ (Bi2201) superconductor by means
of bulk magnetization measurements on high-quality
oxygen-overdoped crystals. In striking contrast with the results
found in the moderately-doped two and three-layer Bi-based
cuprates, Bi2201 exhibits a strong temperature-dependent second-peak effect.
By means of measurements of the in and out-of-plane
first-penetration field we provide direct evidence that this
phenomenon is mainly associated to an increase of the
electromagnetic anisotropy  on warming. The effect of
oxygen-doping $\delta$ on the vortex phase diagram results in both
the irreversibility and second-peak lines shifting to higher
temperatures and fields. This enhanced stability of the Bragg
glass phase suggests that the interlayer coupling between Cu-O
layers increases with $\delta$. In addition, we found that the
critical temperature follows the parabolic relation with the
number of holes per Cu-O plane that holds for most single and
two-layer cuprates.

\end{abstract}



\maketitle

\section*{Introduction}

In the Bi-based series of superconducting cuprates with the
general formula
Bi$_{2}$Sr$_{2}$Ca$_{n-1}$Cu$_{n}$O$_{2n+4+\delta}$ (n=1,2,3), the
single-layer compound Bi$_{2}$Sr$_{2}$CuO$_{6}$ (Bi2201)
\cite{Michel87} is the less assiduously investigated because of
its lower critical temperature $T_{\rm c} \approx 15$\,K and the
difficulty of synthesizing the pure superconducting phase.
However, this compound offers an excellent tool for directly
relating its structural and chemical peculiarities to the
electronic properties of the Cu-O layer. Due to the scarcity of
pure Bi2201 crystals, studies on its vortex phase diagram are
lacking and exhaustive magnetic measurements in the
superconducting state are still to be reported. We present here
the first magnetic bulk measurements on pure Bi2201 crystals, draw
the vortex phase diagram and report its evolution with  oxygen
content $\delta$ in the overdoped regime. In order to do this
study, we have grown high-purity and large crystals of Bi2201
(typical areas of 1 to 5 mm$^{2}$).

In the case of the two and three-layer Bi-based cuprates the
vortex liquid phase spans over a considerable fraction of the
$H-T$ phase diagram \cite{Pastoriza94a, Zeldov95b}. On cooling at
low magnetic fields the vortex matter undergoes a first-order
solidification transition at $T_{\rm m}$
\cite{Pastoriza94a,Zeldov95b}. Upon further cooling the magnetic
response becomes irreversible since pinning sets-in at a
temperature $T_{\rm IL}(H) \lesssim T_{\rm m}(H)$, the so-called
irreversibility line. In the case case of Bi2212 the
low-temperature vortex phase exhibits quasi-crystalline order
\cite{Kim99a,Fasano00a,Fasano05a}. This observation is consistent
with the theoretical proposal that the phase stable at low
temperatures is a Bragg glass \cite{Nattermann90a,Giamarchi94a}.
When increasing field at low temperatures, an order-disorder
transition manifests as the so-called second-peak effect in the
irreversible magnetization \cite{Fisher91a,Khaykovich96a}. This
second-peak effect starts at an onset field, $H_{\rm ON}$, and
presents a local maximum at $H_{\rm SP}$. Recent studies in
several cuprates raised the discussion on identifying the
order-disorder transition field, $H_{\rm OD}$, with either the
onset field $H_{\rm ON}$ \cite{Ravikumar02a,Stamopoulos02} or the
inflection point between $H_{\rm ON}$ and $H_{\rm SP}$ located at
$H_{\rm INF}$ \cite{Giller99a,Nishizaki00a,Radzyner00a,Pissas00a}.
In the case of moderately-doped Bi2212 and Bi2223 the second-peak
maximum $H_{\rm SP}$ is roughly temperature-independent
\cite{Khaykovich96a,Piriou07a,Piriou08}.

 In this work we report that, unexpectedly, in Bi2201 the
second-peak effect  strongly depends on temperature. In order to
elucidate the origin of this phenomenon we performed a detailed
study of the anisotropy parameter, $\gamma = \sqrt{m_{c}/m_{ab}}$
(the ratio between the effective masses along the $c$ axis and
$ab$ plane) and found that it strongly depends on temperature.
Such a dependence is the main responsible for the increase of $H_{\rm SP}(T)$,
 $H_{\rm ON}(T)$ and $H_{\rm INF}(T)$ that we observed in Bi2201 on cooling.

Data on the evolution of the vortex phase diagram with oxygen
doping in Bi2201 crystals was also lacking and is reported here
for the first time. Variations of the doping level greatly modify
the vortex phase diagram of Bi-based cuprates
\cite{Khaykovich96a,Piriou07a,Piriou08,Kishio94a&others,Correa01a}
mainly by inducing changes in the Cu-O interlayer coupling. In
particular, both in Bi2212 and Bi2223, the low-field phase spans
up to higher temperatures and fields on increasing $\delta$. This
is consistent with the measured decrease of $\gamma$ (increase of
interlayer coupling) with oxygen concentration
\cite{Khaykovich96a,Piriou08,Kishio94a&others,Correa01a,Kishio96a}.
We found that in Bi2201 the doping-evolution of the
irreversibility and second-peak lines are qualitatively similar to
those of Bi2212 and Bi2223, suggesting an enhancement of interlayer
coupling on increasing $\delta$.

In the case of Bi2212 and Bi2223 the doping-evolution of the
superconducting parameters has been thoroughly tracked, spanning
from the underdoped (UD) to the overdoped (OD) regime. For the
two-layer compound $T_{\rm c}$ follows a parabolic trend with
carrier density \cite{Allgeier90a,Presland91a,Zhao98a}. The same
law is not fulfilled in the three-layer compound presumably due to
differences in the doping level of the inner and outer Cu-O layers
\cite{Piriou08}. Experimental data on tuning the doping level in
pure Bi2201 is in short supply and controversial. The difficulties
in synthesizing pure Bi2201 have fostered the study of the more
easily processed La and /or Pb-doped Bi2201
\cite{Tarascon90,Schlögl93,Khasanova95,Zhang01a,Eisaki04}.
Presently, the eye has turned back to the pure Bi2201 phase,
however only a single work concerning polycrystalline samples
reports on the dependence of $T_{\rm c}$ on doping \cite{Jean03}.
Sizeable crystals of Bi2201 have been recently grown
\cite{Liang01,Luo07}, but their transition temperatures were not
greater than 8\,K and the effect of post-annealing treatments on
$T_{\rm c}$ was not clarified \cite{Luo07}. We have been able to
tune the doping level over the whole overdoped regime and to
achieve a $T_{\rm c, max}$ = 15\,K, close to the maximum of
16.5 \,K reported for polycrystalline samples \cite{Jean03}.
Furthermore, we report that in our Bi2201 crystals $T_{\rm c}$
follows a parabolic trend with $\delta$, as in the case of several
single and two-layer cuprates \cite{Allgeier90a,Presland91a}.

\section*{Crystal growth and oxygen-doping}

\begin{figure}
\begin{center}
\includegraphics[angle = -90, width=\columnwidth]{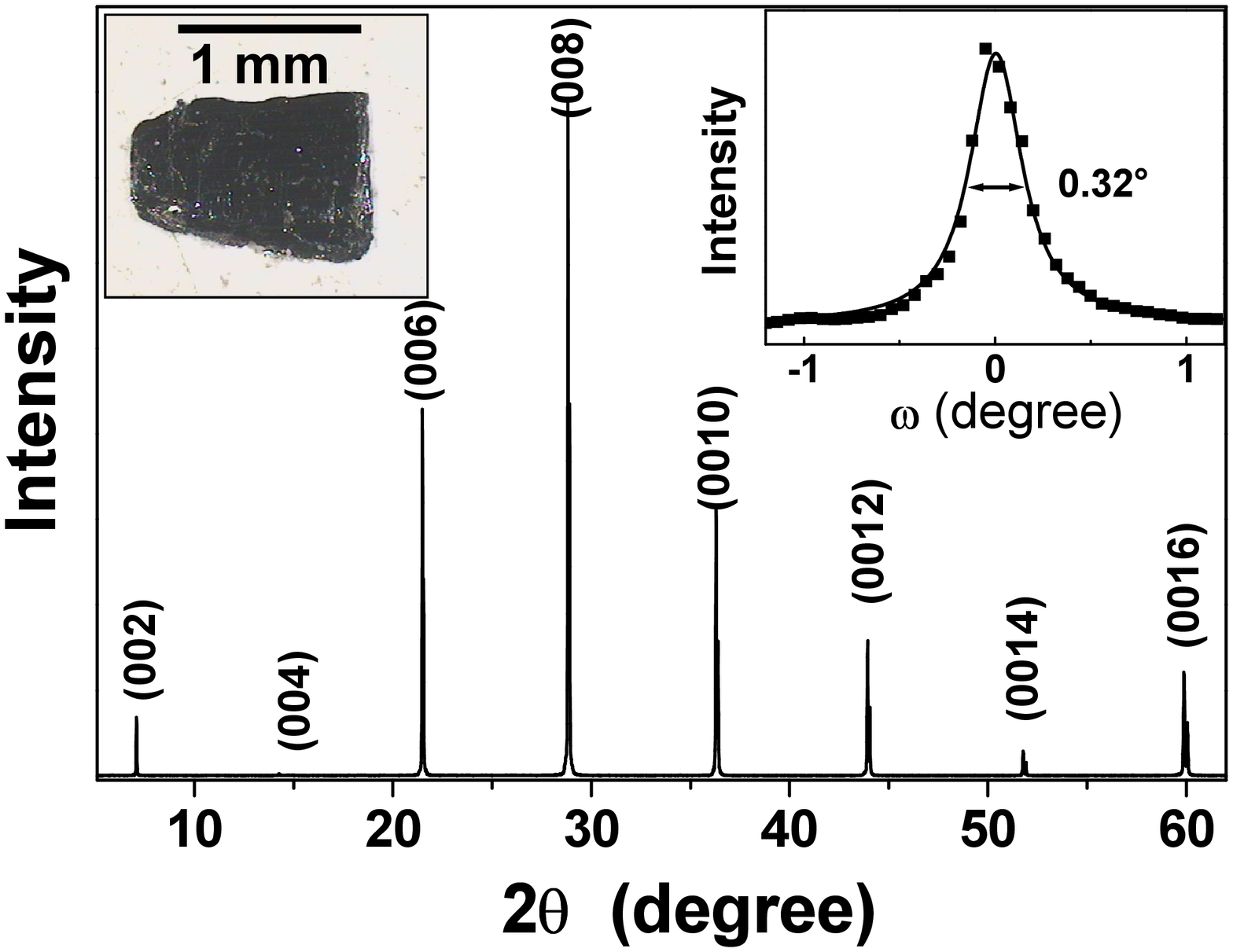}
\caption{X-ray diffraction pattern of one of our typical Bi2201
crystals oriented with the $c$-axis parallel to the scattering
vector. Left-hand insert: picture of the crystal. Right-hand
insert: rocking-curve  exhibiting a full-width at half maximum of
$0.32^\circ$. \label{fig:figure1}}
\end{center}
\end{figure}

Pure Bi2201 crystals were grown by means of the travelling-solvent
floating-zone (TSFZ) method in a home-made two-mirror furnace.
Details on the furnace and the growth technique are described in a
previous work \cite{Giannini04a}. In the case of Bi2201 the growth
of crystals is favored by starting from an excess of Bi in the
nominal composition and by melting in a pure oxygen atmosphere, as
previously reported by other authors \cite{Liang01,Luo07}. The
crystals used in our study were grown from a precursor of nominal
composition Bi$_{2.05}$Sr$_{1.95}$CuO$_{6.025}$. High-purity
Bi$_{2}$O$_{3}$ (99,999\%), SrCO$_{3}$ (99,999\%) and CuO
(99,999\%) were mixed, milled, and calcined at $780-800^\circ$C
during 100-120\,hours in total, with four intermediate manual
grindings. The precursor (feed) rod was cold-pressed in a
cylindrical mold of about 80 mm in length and 7 mm in diameter
and heat-treated in air at $850^\circ$C for 36\,hours. After a
first-zone melting at high travelling-velocity (25\,mm/hour),
performed with the aim of increasing and homogenizing the density
of the feed rod, the slow TSFZ was performed at 0.55\,mm/hour
under an oxygen overpressure of 2$\,$bar. A crystallized end of a
previous sample with the same composition was used as a seed.
Crystals with typical lengths of 1-5\,mm and thicknesses of
0.1-0.2\,mm (see insert of Fig.\,\ref{fig:figure1}) were cleaved
from the core of the crystallized rod.

As-grown crystals are superconducting with an onset of the
$\chi'(T)$ transition at 10\,K and a transition width of  about
 4\,K. The quality of the crystals was checked by
x-ray diffraction (XRD) and energy-dispersive x-ray microprobe
(EDX). The XRD pattern measured in a Bragg-Brentano
$\theta-2\theta$ geometry using a $Cu-K\alpha$ radiation is shown
in Fig.\,\ref{fig:figure1} ($K\alpha_{1}=1.5406$\,\AA,
$K\alpha_{2}=1.5444$\,\AA, $\alpha_{2}/\alpha_{1}=0.5$). In this
configuration, only the $[00l]$ planes contribute to the pattern.
The rocking curve of the $[006]$ reflection is shown in the
right-insert of Fig.\,\ref{fig:figure1}: the full-width at half
maximum is of $0.32 ^\circ$ whereas that of the $I(2\theta)$ peaks
is typically of the order of $0.05 ^\circ$. This XRD data proves
the high crystallinity of our samples. The chemical composition of
the crystals was checked by EDX using a Noran Pioneer X-ray
detector mounted in a Cambridge 438VP
scanning-electron-microscope. The average composition measured
over large crystal areas is
Bi$_{2.05}$Sr$_{1.98}$Cu$_{0.98}$O$_{6.04}$ with errors on the
local deviations in formula units of $\Delta$(Bi) = 0.05,
$\Delta$(Sr) = 0.05 and $\Delta$(Cu) = 0.02.

\begin{figure}
\begin{center}
\includegraphics[angle = -90, width=\columnwidth]{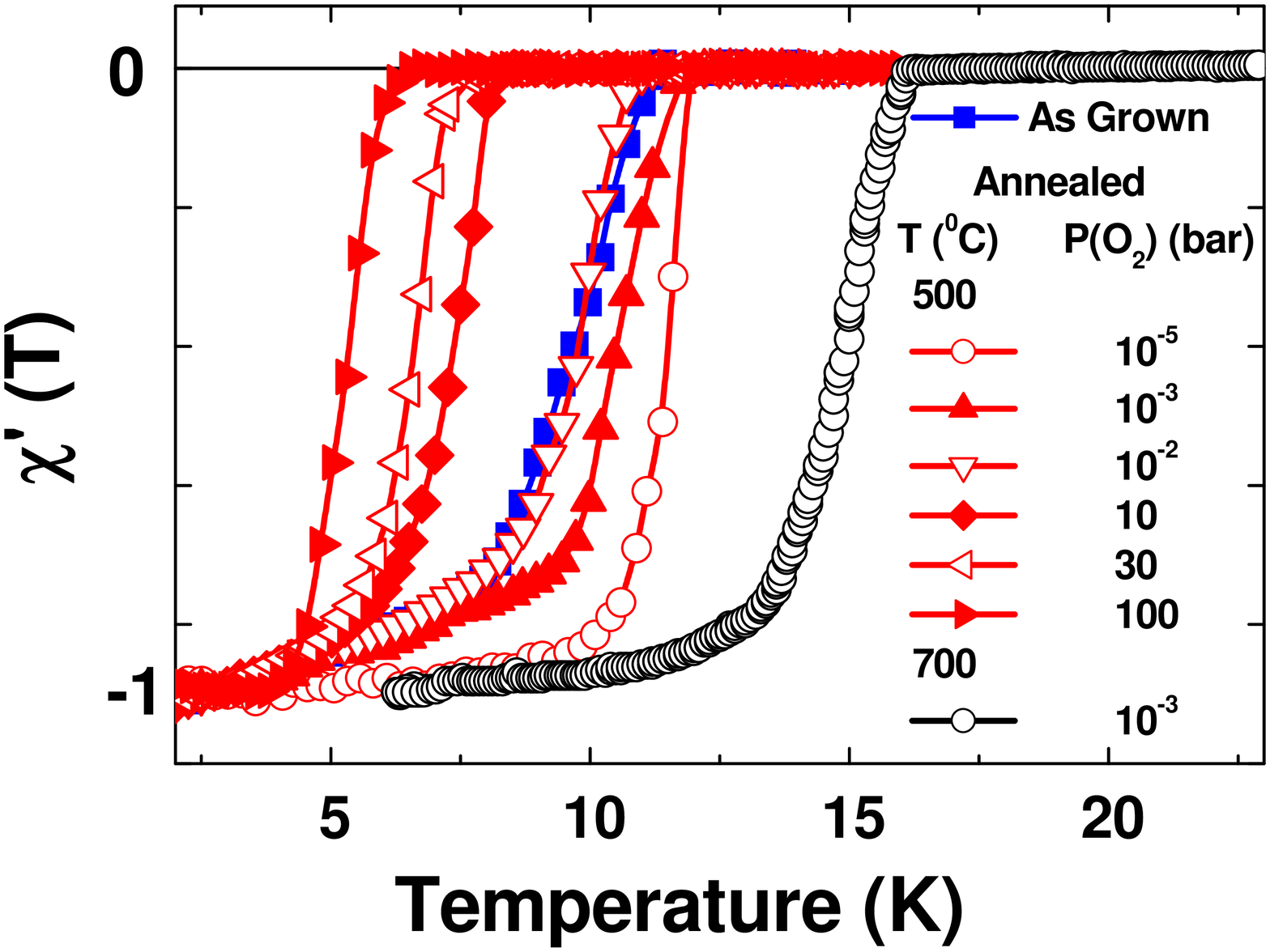}
\caption{Real part of the magnetic susceptibility as a function of
temperature for our samples of Bi2201 annealed under different
oxygen partial-pressures $P(O_{2})$. The annealing treatments were
performed at 500$^{\circ}$C but one at 700$^{\circ}$C. The
$\chi^{'}(T)$ measurements were done with an AC field of 0.1\,Oe
in magnitude and 970\,Hz in frequency. \label{fig:figure2}}
\end{center}
\end{figure}

\begin{figure}
\begin{center}
\includegraphics[width=\columnwidth]{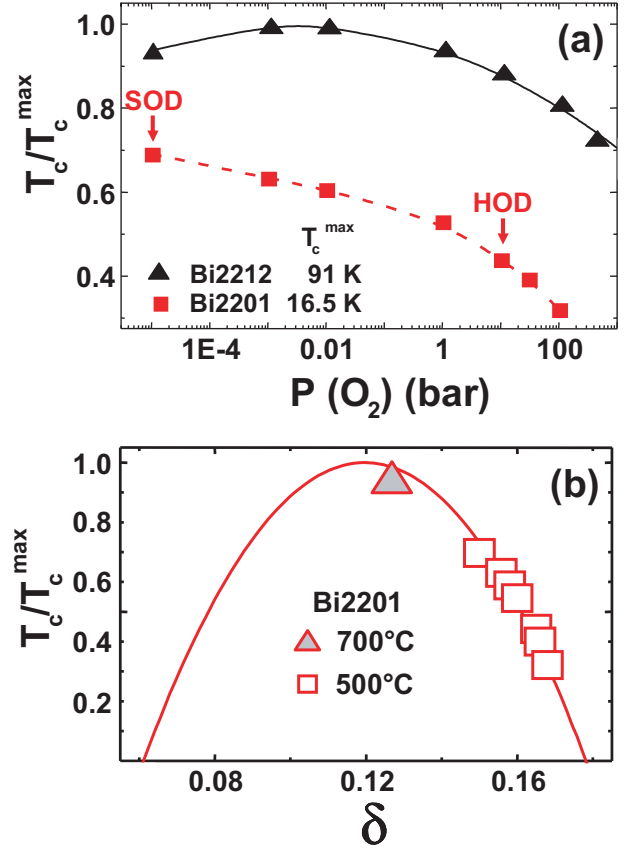}
\caption{(a) Normalized critical temperature of our Bi2201 and
Bi2212 crystals as a function of the annealing oxygen
partial-pressure $P(O_{2})$. The annealing treatments were
performed at 500$^{\circ}$C. In the case of Bi2201 the dotted line
is a guide to the eye whereas for Bi2212 the red line is a fit
with the relation $T_{\rm c}= T_{\rm c}^{max}[1 - 82.6(p -
0.27)^2]$ \cite{Allgeier90a,Presland91a}, where $p = 0.011
\ln{P(O_{2})} + 0.3$ \cite{Zhao98a} is the number of carriers per
Cu-O plane. The two doping regimes considered for the study of the
doping-evolution of the vortex phase diagram are indicated as
slightly(SOD) and highly(HOD)-overdoped. (b) Evolution of $T_{\rm
c}/T_{\rm c}^{max}$ with oxygen content in our Bi2201 crystals.
The doping level was considered as that obtained by means of
thermogravimetric analysis of polycrystalline Bi2201 samples
\cite{Jean03} having the same $T_{\rm c}$ as our crystals. The
line is a fit to the data with the relation  $T_{\rm c}= T_{\rm
c}^{max}[1 - 82.6(p - p_{OPT})^2]$ yielding $T_{\rm
c}^{max}=16.5$\,K, $p = 1.9 \delta$ and $p_{OPT}= 0.23$. A
particular annealing treatment performed at 700$^{\circ}$C (gray
triangle) increased the critical temperature up to 15\,K, a value
close to the maximum of 16.5\,K attributed to optimal doping in
Ref.\,\onlinecite{Jean03}. \label{fig:figure3}}
\end{center}
\end{figure}

 In order to tune and homogenize the carrier concentration the
crystals were annealed during 24 to 48\,hours at 500$^{\circ}$C,
under various oxygen partial-pressures $P(O_{2})$.  Annealing
treatments longer than 48\,hours did not affect either the
transition temperature $T_{\rm c}$ or the transition width $\Delta
T_{\rm c}$. Magnetic susceptibility measurements reveal single and
relatively sharp superconducting transitions with widths ranging
from 1 to 3.5\,K. Examples of superconducting transitions for
various doping levels are shown in Fig.\,\ref{fig:figure2}.
$T_{\rm c}$ is considered as the temperature at which the
temperature-derivative of the AC susceptibility, $\partial
\chi^{'} /\partial T$, and DC magnetization, $\partial M /\partial
T$, are peaked. Figure\,\ref{fig:figure3} shows the dependence of
$T_{\rm c}/T_{\rm c}^{max}$ on $P(O_{2})$ for our Bi2201 crystals.
Each point corresponds to at least 5 samples with the same
critical temperature within the error. Superconductivity is
suppressed when annealing at $P(O_{2}) = 400$\,bar. As expected
\cite{Jean03}, after  annealing at 500$^{\circ}$C Bi2201 is in the
overdoped regime and still remains overdoped at any annealing
pressure down to 10$^{-5}$ bar.

It is important to point out that in the case of polycrystalline
samples the maximum $T_{\rm c}$ value of 16.5\,K is only reached
when annealing at 700$^{\circ}$C \cite{Jean03}. In our crystals, a
treatment at 700$^{\circ}$C and $10^{-3}$\,bar enhances the
critical temperature to 15\,K (see Fig.\,\ref{fig:figure2}).
However, after this annealing the crystals present no mirror-like
surfaces and exhibit broad transitions with widths between 4 and
10\,K. Since the Bi2201 phase is at its stability limit at such
annealing temperature, these samples are likely to have degraded
domains.

Our Bi2201 crystals quantitatively follow  the same $T_{\rm c}$
\textit{vs.} $P(O_{2})$ behavior than polycrystalline samples.
This allows us to consider the doping level of our crystals as
that obtained by means of thermogravimetric analysis of
polycrystalline Bi2201 samples \cite{Jean03} with the same $T_{\rm
c}$. The evolution of  $T_{\rm c}/T_{\rm c}^{max}$ with
hole-doping is presented in Fig.\,\ref{fig:figure3} (b). The data
is very well fitted by a parabolic evolution of $T_{\rm c}$ with
$p$, the number of holes per Cu-O plane, $T_{\rm c}= T_{\rm
c}^{max}[1 - 82.6(p - p_{OPT})^2]$ \cite{Allgeier90a,Presland91a},
yielding $T_{\rm c}^{max}=16.5$\,K and $p - p_{OPT}= 1.9 \delta -
0.23$. Therefore, Bi2201 follows the $T_{\rm c}$ \textit{vs.}
$p$ relation reported to be obeyed by other single and two-layer
cuprates.

\section*{Effect of oxygen-doping on the vortex
phase diagram of B$\textmd{i}$2201}

The results presented in this section were obtained in the same
sample for two different doping levels within the overdoped
regime. The post-annealing treatments were performed at
500\,$^{\circ}$C and at pressures of 10$^{-5}$ and $10$\,bar,
resulting in $T_{\rm c}=(11.4 \pm 0.5)$ and $(8.0 \pm 0.8)$\,K for
the slightly (SOD) and highly-overdoped (HOD) regions,
respectively. The doping level corresponds thus to $\delta = 0.15$
for the SOD and $0.165$ for the HOD regimes (see
Fig.\,\ref{fig:figure3} (b)).

\begin{figure}[ttt]
\includegraphics[ width=\columnwidth]{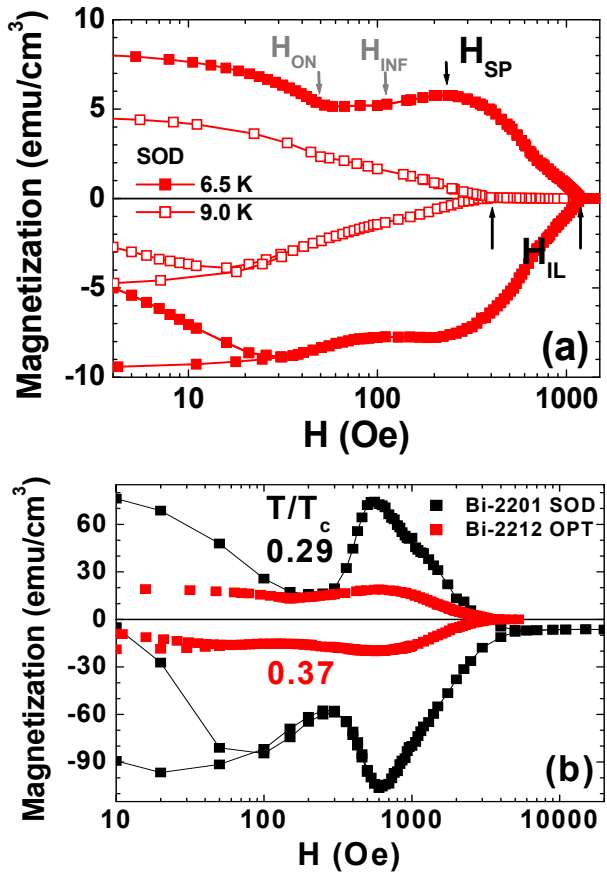}
\caption{(a) Magnetization loops of Bi2201 in the
slightly-overdoped regime (SOD)[$T_{\rm c}=(11.4 \pm 0.5)$\,K].
The arrows indicate the irreversibility, $H_{\rm IL}$, and
characteristic second-peak fields,  $H_{\rm SP}$ (local maximum),
$H_{\rm ON}$ (onset), and $H_{\rm INF}$ (inflection or kink
point). The measurements were performed at a sweep rate of
$10^{-2}$\,Oe/s. (b) Locus of magnetization loops for Bi2201 and
Bi2212 (from Ref.\,\onlinecite{Correa01a}) at comparable reduced
temperatures and doping levels. \label{fig:figure4}}
\end{figure}

The effect of oxygen doping on the vortex phase diagram of Bi2201
was investigated by means of bulk magnetization. The measurements
were performed using a MPMS2 SQUID magnetometer, a PPMS
measurement system and a vibrating-sample magnetometer (VSM).
 In the first set of measurements we focus on the
doping-evolution of the irreversibility,  $H_{\rm IL}$, and
second-peak lines, $H_{\rm SP}$, $H_{\rm ON}$ and $H_{\rm INF}$.
 These lines were obtained from magnetization
\emph{vs.} magnetic field measurements, $M(H)$, and from field
cooled-zero field cooled (FC-ZFC) temperature-dependent
magnetization curves. The magnetic field was applied
parallel to the crystal $c$-axis and swept at rates of
$10^{-2}$\,Oe/s (SQUID and PPMS magnetometers), 1 and 10\,Oe/s (VSM).
 Figure \ref{fig:figure4} (a) shows examples of
magnetization loops for the SOD regime.

The onset of the irreversible magnetic response was estimated as
the field at which the two branches of the magnetization loop
merge, as indicated with the arrows in Fig.\,\ref{fig:figure4}(a).
Estimating the irreversibility temperature from the splitting of
the FC-ZFC branches in $M(T)$ curves yielded similar values. Three
different effects can be at the origin of an irreversible magnetic
response in superconductors: bulk pinning, Bean-Livingston surface
barriers \cite{Bean64b} and geometrical barriers
\cite{Zeldov94a,Indenbom94a}. The Bean-Livingston surface barrier
only produces a significant irreversible behavior in the case of
extremely smooth surfaces \cite{DeGennes}. In real samples with
sharp corners and irregular edges, the effect of this barrier is
of lesser importance. In general, macroscopic magnetization
measurements are not able to ascertain which of the other two
contributions is dominant when measuring an irreversible magnetic
response.  The effect of geometrical barriers can be revealed by
conveniently modifying the sample geometry. In prism-like samples
of Bi2212, the geometrical barriers are suppressed and the
irreversibility line deviates significantly from the melting line
determined by bulk properties \cite{Majer95a}. However, for
platelet-like Bi2212 samples, independent measurements of the
irreversibility and melting lines indicate that both lines
incidentally merge \cite{Pastoriza94a}. Therefore, in this case,
both bulk pinning and geometrical barriers contribute to the
irreversible magnetic response at fields lower than $H_{\rm IL}$.
Studying how $H_{\rm IL}$ deviates from the
melting line when changing the sample geometry in Bi2201 is far beyond
the aim of this work. However, it is
reasonable to assume that in our platelet-like samples the effect
of bulk pinning may become relevant at fields equal to or slightly
lower than $H_{\rm IL}(T)$.

\begin{figure}[ttt]
\includegraphics[angle = -90, width=\columnwidth]{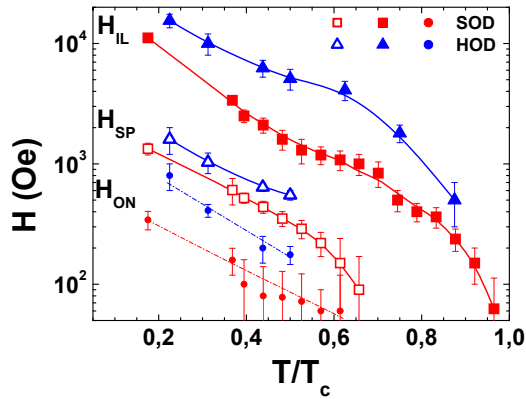}
\caption{ Vortex phase diagram for the same Bi2201 sample in the
slightly ($\Box$ SOD) [$T_{\rm c}=(11.4 \pm 0.5)$\,K] and
highly-overdoped ($\triangle$ HOD ) [$T_{\rm c}=(8.0 \pm 0.8)$\,K]
regimes. The irreversibility line, $H_{\rm IL}(T)$ (full symbols,
full lines), and the maximum, $H_{\rm SP}(T)$ (open symbols, full
lines) and onset lines, $H_{\rm ON}(T)$ (full symbols, dashed
lines), of the second-peak effect are shown. The points extracted
from magnetization loops correspond to measurements performed at a
sweep rate of $10^{-2}$\,Oe/s.  The error bars, when not visible,
are within the size of the symbols. \label{fig:figure5}}
\end{figure}

The effect of oxygen-doping on the irreversibility line is shown
in the $H$ \emph{vs.} $T/T_{\rm c}$ phase diagram of
Fig.\,\ref{fig:figure5}. At any temperature, the irreversibility
field is enhanced with increasing the oxygen content. A similar
evolution of $H_{\rm IL}$ with doping was reported for Bi2212
\cite{Correa01a} and Bi2223  \cite{Piriou07a,Piriou08}. The larger
extent of the irreversible vortex solid in the HOD regime
indicates that the interlayer coupling is enhanced when $\delta$
increases. Therefore, the doping-evolution of $H_{\rm IL}$
suggests a decrease of $\gamma$ when increasing $\delta$ in the
overdoped regime.

The second-peak effect is observed in  $M(H)$ curves as a
peak-valley structure (as also observed in
La-doped Bi2201 \cite{Amano2004}), see for example the
magnetization loops in Fig.\,\ref{fig:figure4}. The characteristic
$H_{\rm SP}$ field associated with the local maximum of the
magnetization is clearly evident. However, since in Bi2201 the
second-peak feature is broad, the onset and inflection points are
poorly resolved. The broad locus of $M(H)$ in Bi2201 is
illustrated in Fig.\,\ref{fig:figure4}(b) by comparing with data
obtained in Bi2212 at similar reduced temperatures and doping
levels \cite{Correa01a}. Our bulk magnetization measurements on Bi2201
detect the second-peak effect within a temperature range
$0.18 \leq T/T_{\rm c} \leq 0.65$. One should notice that
the $H_{\rm SP}$ line seems to end below the $H_{\rm IL}$ line.
High-resolution local magnetic measurements allowed to study how
the $H_{\rm SP}$ line terminates at low fields and high temperatures in the case of YBCO
\cite{Stamopoulos02b}. Our bulk measurements do not enable us
 to infer anything about the end point of the $H_{\rm SP}$ line.

\begin{figure}[ttt]
\includegraphics[angle = -90, width=1.1\columnwidth]{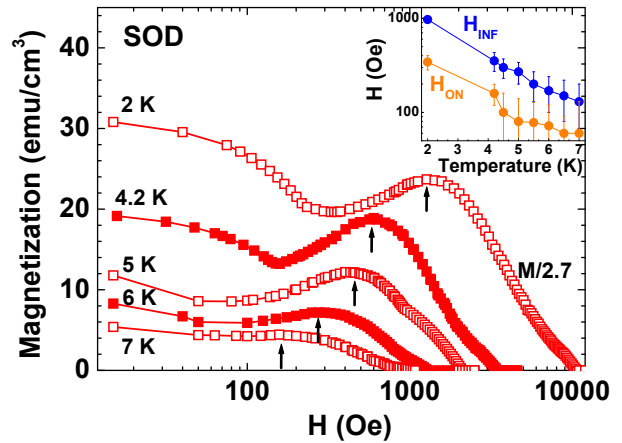}
\caption{Upper-right quadrant of the magnetization loops $M(H)$
measured in the SOD regime of Bi2201 [ $T_{\rm c}$=(11.4 $\pm$ 0.5
)\,K ] at different temperatures. The 2\,K curve is
normalized by a 2.7 factor in order to include all curves in the
same scale. A temperature-dependent second-peak
effect is detected between 2 and 7\,K ($0.18 \leq T/T_{\rm
c}^{max} \leq 0.65$). Arrows indicate the local maxima of the
magnetization $H_{\rm SP}$.
Insert: Temperature-evolution of the onset, $H_{\rm
ON}$, and inflection, $H_{\rm INF}$, characteristic fields of the second-peak
feature. Only data up to 7\,K are shown since these two fields are
not clearly resolved at higher temperatures. The measurements were
performed at a sweep rate of $10^{-2}$\,Oe/s.
\label{fig:figure6} }
\end{figure}

\begin{figure}[ttt]
\includegraphics[angle = -90, width=\columnwidth]{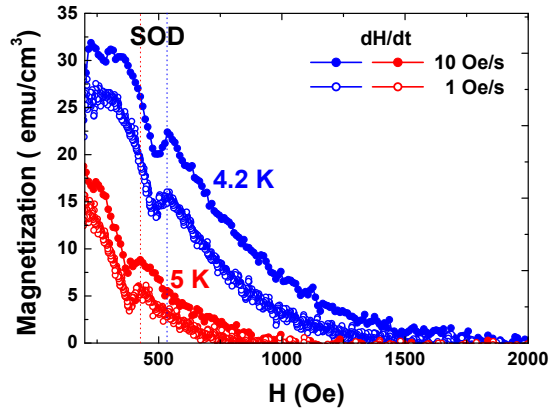}
\caption{Upper-right quadrant of the magnetization loops $M(H)$
measured in the SOD regime of Bi2201 [ $T_{\rm c}$=(11.4 $\pm$ 0.5
)\,K ] at two different sweep rates and temperatures. The field
location of the maximum of the second-peak effect is independent
of the sweep rate and decreases on warming.
\label{fig:figure7} }
\end{figure}

As well documented in the literature, the locus of the second-peak
effect and eventually its detection might have a dependence on the
electric-field level influenced by the magnetic-field sweep-rate.
For this reason we have also studied the effect of the sweep rate
on the field-location of the characteristic field that we can
track with low error, $H_{\rm SP}$. The magnetization loops shown
in Fig.\,\ref{fig:figure7} were measured on a slightly-overdoped sample
(from the same batch and of the same $T_{\rm c}$ as the one presented in Fig.\,\ref{fig:figure6})
at faster sweep rates than that used for the curves presented in Fig.\,\ref{fig:figure6} ($10^{-2}$ Oe/s).
The right-upper quadrant of the loop is shown for two temperatures,
4.2 and 5 K, and two sweep rates, 10 Oe/s and 1 Oe/s. It is clear
from Figs.\,\ref{fig:figure6} and \ref{fig:figure7}, that $H_{\rm SP}$ does not
depend on the sweep-rate. The differences in the shape of the peak as well as in the apparent $H_{\rm IL}$ of Fig.\,\ref{fig:figure7},
as compared to Fig.\,\ref{fig:figure6}, can be due to a lesser doping homogeneity in the larger sample used for the VSM experiment.

 The magnetization curves presented in
Figs.\,\ref{fig:figure6} and \ref{fig:figure7} indicate an
unexpected result for a moderately-doped Bi-based cuprate: unlike
the two and three-layer compounds
\cite{Khaykovich96a,Piriou07a,Piriou08}, in Bi2201 $H_{\rm SP}$
decreases by one order of magnitude on warming. Both fields
$H_{\rm ON}$ and $H_{\rm INF}$ follow a similar trend as shown in
the insert of Fig.\,\ref{fig:figure6}. Because of the still unsolved
controversy about which is the true signature of the
order-disorder transition
\cite{Ravikumar02a,Stamopoulos02,Giller99a,Nishizaki00a,Radzyner00a,Pissas00a},
we will rather discuss the temperature dependence of $H_{\rm SP}$
that we can determine better than $H_{\rm ON}$ and $H_{\rm INF}$.

The second-peak effect is associated with the vortex-solid
order-disorder phase transition at which the elastic energy equals
the pinning energy
\cite{Giamarchi94a,Ertas96a&Giamarchi97a&Vinokur98a&Kierfeld98a}.
These two energy terms depend on the penetration depth, coherence
length and anisotropy of the material, as well as on the
temperature-dependent pinning parameter \cite{Giamarchi94a}.
Considering the two-fluid model \cite{Blatter94a}, $\lambda_{\rm
ab}$(T) and $\xi_{\rm ab}$(T) vary by only $3$\,\% within the
temperature range in which the second-peak effect is detected in
Bi2201. Such a small variation can not account for the observed
temperature-dependent H$_{\rm SP}$.  According to theoretical
predictions \cite{Giamarchi94a}, either a small and/or
temperature-dependent anisotropy parameter, or an important
temperature-dependent pinning parameter, or both, can produce a
non-constant $H_{\rm SP}$(T).

A temperature-dependent second-peak effect has been reported in
several cuprates
\cite{Ooi96,Giller97,Ooi98a,Mumtaz98,Giller99a,Baziljevich00,
Stamopoulos02,Darminto02a,Chowdhury04}. Roughly, two types of
temperature-evolution for $H_{\rm SP}$ are observed. In one class
of materials $H_{\rm SP}$ decreases non-monotonically on warming,
presenting a valley structure at low temperatures. In this first
group, including strongly-doped Bi2212 samples
\cite{Khaykovich96a,Ooi96,Ooi98a,Mumtaz98,Baziljevich00,Darminto02a},
the temperature-dependent $H_{\rm SP}$ is associated to enhanced
disorder with respect to optimally-doped pure samples. The second
class displays a monotonous decrease of the second-peak field on
warming, distinctly detected up to temperatures very close to
$T_{\rm c}$. Notorious examples of cuprates belonging to this
group are YBa$_2$Cu$_3$O$_{7 - \delta}$ \cite{Giller99a},
Nd$_{1.85}$Ce$_{0.15}$CuO$_{4 - \delta}$ \cite{Giller97},
HgBa$_2$CuO$_{4 + \delta}$ \cite{Stamopoulos02} and
TlBa$_2$CuO$_{6}$ \cite{Chowdhury04}. In this case the
temperature-dependent second-peak effect is associated with the
combination of a relatively low anisotropy, a decrease of the
pinning energy on warming, and the significant increase of
$\lambda$ and $\xi$ close to $T_{\rm c}$.

Since in our pure Bi2201 samples $H_{\rm SP}$  monotonously
decreases on warming, the role of disorder is likely to be
moderate and this temperature dependence would then be ascribed to
a temperature-dependent anisotropy.  In order to quantitatively
study this issue, the theoretical approach that describes the
second-peak effect as the manifestation of an order-disorder phase
transition should be considered
\cite{Giamarchi94a,Ertas96a&Giamarchi97a&Vinokur98a&Kierfeld98a}.
However, in our case two reasons hindered us to apply this
approach. First, no data on the magnitude of the anisotropy
parameter in Bi2201 was previously available in the literature.
Second, since it is still controversial which field is the fingerprint
of the order-disorder transition \cite{Ravikumar02a,Stamopoulos02,Giller99a,Nishizaki00a,Radzyner00a,Pissas00a},
\emph{local}-magnetization and/or partial
magnetization loops data are mandatory. This analysis is beyond
the aim of this work. Instead, we directly measured the magnitude
of the anisotropy parameter and study its role in the
temperature-dependent $H_{\rm SP}$.

The anisotropy parameter was estimated from directional
measurements of the first-penetration field, as previously
reported for the Bi2223 compound \cite{Piriou08}. Within the
London approximation $\gamma = H_{\rm c1}^{\perp}/H_{\rm
c1}^{\parallel}$ \cite{Blatter94a}, with $\perp$ and $\parallel$
meaning perpendicular and parallel to the $ab$ plane. The sample
was aligned in both configurations using a home-made rotation
system that reduces the misalignment uncertainty to $\sim
0.5^{\circ}$. In both configurations the first penetration field
$H_{\rm p}$ was considered as that at which the magnetization
shows a detectable relaxation, associated with the entrance of the
first vortex. This was done by measuring at every field the
relaxation of the magnetic moment during 1\,hour. This method
reduces the effect of surface and geometrical pinning barriers
\cite{Niderost98a} and is not affected by the error in identifying
$H_{\rm p}$ from the deviation from linearity in $M(H)$. We assume
that in our experiment $H_{\rm p}$ corresponds to the lower
critical field $H_{\rm c1}$ that is borne out by the absence of
any asymmetries in the low-temperature hysteresis loops. This
confirms that surface pinning effects are negligible. The effect
of demagnetizing factors was corrected considering the Meissner
slope for both configurations. The large demagnetizing effects
strongly reduces the difference between $H_{\rm p}$ and $H_{\rm
c1}$, thus reducing the uncertainty in measuring the true critical
field \cite{Burlachkov93a}. In our case, the estimation of the
anisotropy parameter from directional measurements of $H_{\rm c1}$
is preferable to that obtained from $H_{\rm c2}$ measurements
\cite{Niderost98a}. The latter are only possible at high fields
and/or temperatures,  far from the region over which we measured
the temperature dependence of $H_{\rm SP}$.

\begin{figure}[ttt]
\includegraphics[angle = -90, width=\columnwidth]{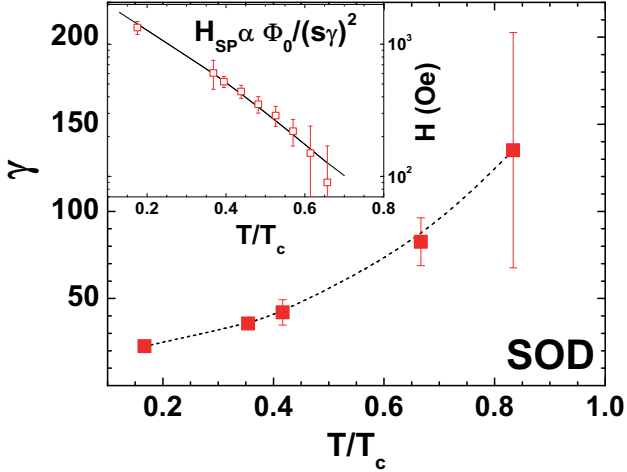}
\caption{Temperature-evolution of the anisotropy parameter for
slightly-overdoped (SOD) Bi2201 [$T_{\rm c}$ = (11.4 $\pm$ 0.5)
\,K]. Insert: Fit of the second-peak data with the
relation $H_{\rm SP}(T) \propto \Phi_{0}/(s \gamma(T))^2$ with
$\gamma(T)$ estimated from the interpolation of the data in the main panel.
\label{fig:figure8} }
\end{figure}

Figure \,\ref{fig:figure8} shows that for SOD Bi2201 $\gamma$ is
strongly temperature-dependent: it increases from roughly 25 to 80
in the temperature-range in which $H_{\rm SP}$ is detected and
further increases up to 130 at $T/T_{\rm c} \sim 0.8$. The data
also indicates that Bi2201 presents a moderate-to-high
electromagnetic anisotropy in the SOD regime. At low temperatures
the anisotropy parameter of SOD Bi2201 is intermediate between
that of SOD Bi2223 ($\sim 20$ \cite{Piriou08}) and SOD Bi2212
($\sim 100$ \cite{Correa01b}). Local magnetic measurements in
moderately-doped Bi2212 revealed that $\gamma$ increases on
warming in the $T/T_{\rm c}$ range $0.74-0.96$
\cite{Konczykowski06a}.

 In the case of Josephson-dominated coupling between the Cu-O planes, $s \gamma <
\lambda_{ab}$, the order-disorder transition field is inversely
proportional to the anisotropy \cite{Koshelev98a}. Within the
temperature range in which $H_{\rm SP}$ is detected in SOD
Bi2201, $s\gamma < \lambda_{ab}= (3200 \pm 200)$\,\AA, with
$s=12.3$\,\AA\,\cite{Vedeneev99} the distance between adjacent
Cu-O planes. The penetration depth $\lambda_{ab}$ was obtained by
fitting the temperature-dependent $H_{\rm c1}^{\perp}$ within the
London model, considering the two-fluid expression of
$\lambda_{ab}(T)$ and $\xi_{ab}(T)$ \cite{xi}. Therefore, since in
the considered temperature range the Josephson coupling is
dominant, the relation $H_{\rm SP} \propto \Phi_{0}/(s \gamma)^2$
should be valid. The insert of Fig.\,\ref{fig:figure8} shows
the excellent agreement between the $H_{\rm SP}$ data and $\propto
\Phi_{0}/(s \gamma(T))^2$. This
finding constitutes a strong evidence that in Bi2201 $H_{\rm SP}(T)$
is governed by the temperature-dependence of the anisotropy and
that the pinning parameter is moderate.

This last statement is further supported by critical current data.
A less relevant role of disorder implies a reduction on the
critical-current density-ratio, $J_{c}/J_{0}$, with $J_{\rm 0}= 4c
\Phi_{\rm 0}/ 12 \sqrt{3} \pi \lambda^{2}_{\rm ab} \xi_{\rm ab}$
the depairing-current density. The lower magnitude of $J_{c}$ in
Bi2201, compared to that of Bi2212, is already suggested by the
magnetization loops shown in Fig.\,\ref{fig:figure4}(b). In
addition, in Bi2212 a temperature-dependent $H_{\rm SP}$,
attributed to enhanced disorder \cite{Ooi98a, Darminto02a}, was
only observed for the extremely-overdoped regime in which the
critical current is significantly larger than in the
optimally-doped regime \cite{Correa01b}.

\begin{figure}[ttt]
\includegraphics[angle = -90, width=\columnwidth]{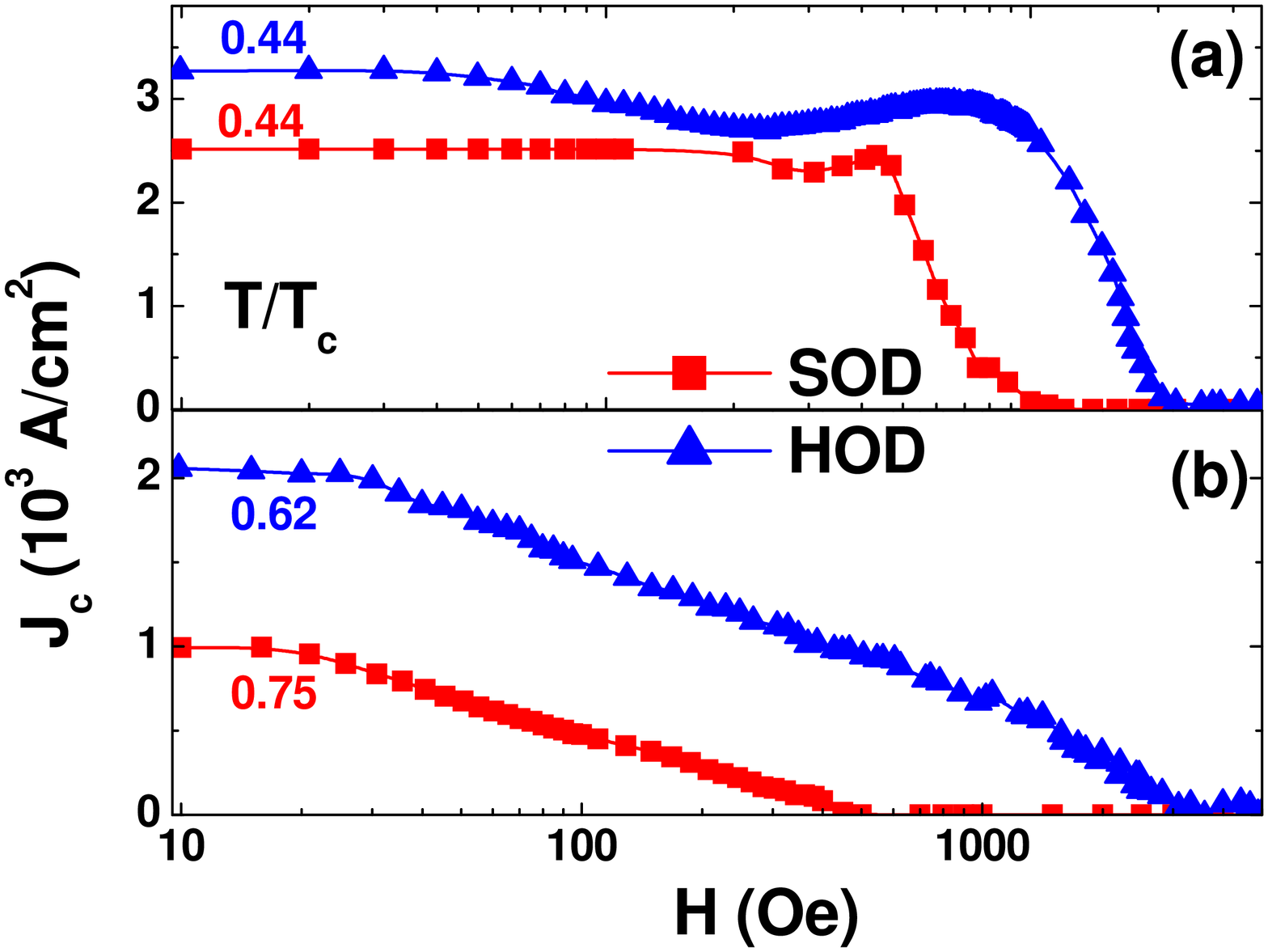}
\caption{Critical current density as a function of magnetic field
, $J_{c}(H)$, for  slightly (SOD) and highly-overdoped (HOD)
Bi2201 obtained from the width of magnetization loops. (a) In the
low-temperature data the second-peak is clearly observed whereas
(b) in the high-temperature data  $H_{\rm SP}$ is no longer
resolved. \label{fig:figure9} }
\end{figure}

The critical current of our Bi2201 crystals is obtained from
magnetization loops measured at different temperatures.
Considering the Bean model \cite{Bean64a}, at a given temperature
$J_{\rm c} (T,H) \sim (3c/2R) \Delta M (T,H)$, where $\Delta M
(T,H)$ is the separation between the two branches of the
magnetization loop at a field $H$, $c$ is the speed of light and
$R$ is the radius of an equivalent cylindrical sample
\cite{Bean64a}. Figure\,\ref{fig:figure9} shows examples of
$J_{c}$ curves for SOD and HOD Bi2201 at low and
high-temperatures. At low temperatures the second-peak effect is
clearly observed at intermediate fields and $J_{c}$ is found to be
field-independent at low fields. This suggests that at low
temperatures and fields the vortex lines are individually pinned,
as also observed in Bi2212 \cite{Correa01c} and Bi2223
\cite{Piriou08}. At high temperatures the second-peak effect is no
longer resolved and the critical current is strongly
field-dependent. These findings are observed in both the SOD and
HOD regime. Within the temperature-range studied, the critical
current density in the HOD is larger than in the SOD regime.

The critical current curves presented in  Fig.\,\ref{fig:figure9}
allow us to estimate that the ratio $J_{c}/J_{0}$  in
moderately-overdoped Bi2201 ($J_{0}(T/T_{\rm c}=0.3) \sim
10^{3}$\,A/cm$^{2}$) is 1-2 orders of magnitude smaller than that
of extremely-overdoped Bi2212 \cite{Correa01b} at similar
reduced-temperatures and fields. Thus the role of disorder in our
overdoped Bi2201 samples is much less relevant than in the Bi2212
samples presenting a temperature-dependent $H_{\rm SP}$
\cite{Ooi98a, Darminto02a}. This evidence strengthen our argument
that in Bi2201 the observed temperature-dependent second-peak
effect is mainly the consequence of an enhancement of anisotropy
on warming.

Finally, we would like to discuss another important result evident
from the vortex phase diagram presented in
Fig.\,\ref{fig:figure5}: $H_{\rm SP}(T)$  shifts towards higher
fields on increasing doping. The same qualitative behavior was
reported in Bi2212
\cite{Khaykovich96a,Kishio94a&others,Correa01a}, Bi2223
\cite{Piriou07a,Piriou08}  and other cuprates
\cite{Hardy94a&Shibata02a&Masui04a}. This evolution of $H_{\rm
SP}(T)$ is consistent with an enhancement of coupling between the
Cu-O planes with increasing oxygen concentration, as also
suggested by the doping dependence of $H_{\rm IL}$. Since we
showed that in Bi2201 $H_{\rm SP}$ is inversely proportional to
$\gamma^2$, the doping evolution of $H_{\rm SP}(T)$  allows the
estimation of the anisotropy in the HOD regime. Considering the
data of Fig.\,\ref{fig:figure5} we estimate a 15\% decrease of
$\gamma$ for the HOD with respect to the SOD regime.

\section*{Conclusions}

We have grown pure and large Bi2201 single crystals and tuned the
doping level over the whole overdoped regime. The critical
temperature of Bi2201 follows a parabolic dependence with the
number of holes per Cu-O plane, as found in several single and
two-layer cuprates.

The doping-evolution of the vortex phase diagram was studied by
means of bulk magnetic measurements. Varying the oxygen concentration
affects the vortex phase diagram  in a way that is consistent with
an enhancement of the coupling between Cu-O layers with increasing
$\delta$. This result is in agreement with data reported for the
two and three-layer compounds. However, in striking
contrast with results found  in those compounds, Bi2201 presents a
strong temperature-dependent second-peak effect. The electronic
anisotropy increases on warming and $H_{\rm SP}$ scales with
$1/\gamma^{2}$, as expected for Josephson-dominated interlayer
coupling. Since in addition the relevance of pinning in Bi2201 is
smaller than in the other two Bi-based cuprates,  we conclude that
the temperature-dependent $H_{\rm SP}$ can be mainly ascribed to
the temperature evolution of the anisotropy.

\vspace{1cm}

This work was supported by the Swiss National Science Foundation,
(Project 200020-118029). The authors acknowledge M. Konczykowski
for stimulating discussions and V. Correa for providing us the
software version of his published data in Bi2212 \cite{Correa01a}.


\begin{references}
\bibitem{xi} In order to calculate $\xi_{ab}(T)$ we have taken $\xi_{ab}(0) = 40$\,\AA\ according to Ref.\,\onlinecite{Vedeneev99}.
\end{references}
\end{document}